\RequirePackage{fix-cm}
\documentclass[smallextended]{svjour3}       % onecolumn (second format)
\smartqed  % flush right qed marks, e.g. at end of proof
\usepackage{amsmath}
\usepackage{xcolor,graphicx,float}
\usepackage{bbding}
\usepackage{mathrsfs}
\usepackage[justification=centering]{caption}

\begin{document}

\title{Quantum security computation on shared secrets %\thanks{Grants or other notes
}
\author{Hai-Yan Bai  \and
         Zhi-Hui Li  \and
         Na Hao %etc.
}


\institute{Hai-Yan Bai, Zhi-Hui Li\Envelope ,Na Hao \at
             School of Mathematics and Information Science, Shaanxi Normal University, Xi'an 710119, China \\
              \email{ lizhihui@snnu.edu.cn}           %  \\
           \and
}

\date{Received: date / Accepted: date}
\maketitle
\begin{abstract}
Ouyang et al. proposed an $(n,n)$ threshold quantum secret sharing scheme, where the number of participants is limited to $n=4k+1,k\in Z^+$, and the security evaluation of the scheme was carried out accordingly. In this paper, we propose an $(n,n)$ threshold quantum secret sharing scheme for the number of participants $n$ in any case ( $n\in Z^+$ ). The scheme is based on a quantum circuit, which consists of Clifford group gates and Toffoli gate. We study the properties of the quantum circuit in this paper and use the quantum circuit to analyze the security of the scheme for dishonest participants.
\end{abstract}

\section{Introduction}
\label{intro}
Quantum computing is a new type of computing mode that regulates quantum information and follows the laws of quantum mechanics. That is, it enables quantum bits to achieve the purpose of programming in different quantum logic gates sequentially, and all kinds of quantum algorithms can be realized by combining different quantum logic gates. From the perspective of computational efficiency, because of the existence of quantum mechanical superposition, some quantum algorithms are faster than the conventional general computer when solving problems.
\par In the field of quantum cryptography, aside from the quantum key distribution[1,2], the quantum computation of security computing has attracted the attention of people, such as secure multiparty computation [3], blind computation [4-7] and verifiable delegated computation [8-12]. The earliest quantum algorithm is proposed by Jozsa and Deutsch. This quantum algorithm shows the computing power that classical computers do not have [13]. Ouyang et al. give a different way of security computing quantum algorithm [14], and the security evaluation of quantum secret sharing circuit.
\par In 1999, Hillery [15] and others proposed a quantum secret sharing scheme firstly. The emergence of this concept provides an effective way for the security of quantum state secrets. After that, the quantum secret sharing scheme has received rapid attention and development [16-22]. The so-called $(k,n)$-threshold quantum secret sharing scheme will mean the quantum state secret distributed to $n$ participants such that no group fewer than $k$ participants can reconstruct the secret quantum state [23-25], and any $k$ participants can reconstruct the secret quantum state.
\par In the quantum circuits, the implementation of the unitary transformation of the quantum state is equivalent to the role of the logic gate in the quantum state. In any dimension Hilbert space, if the unitary transformation of the corresponding quantum state can be realized through the combination of elements in a logic gate group, we call such a logic gate group a universal logic gate group (universal gate), hereinafter referred to as universal gate. It has been proved that Toffoli gates can realize all unitary transformation [26] in any dimension Hilbert space, but it is a 3 qubits quantum gate, which is difficult from the perspective of physical realization. However, in any case, the universal gates of quantum computing, like the classical logic gate group, have different forms of expression, as the Ref.[27] points out that any of the $T$ gate, the controlled phase gate and the Toffoli  gate and Clifford group gates can form a group of universal gate. In this regard, Ouyang et al. considers the universal gate composed of discrete Clifford group gates and $T$ gate [14].
\par In this paper, we consider the situation that Clifford group gates and Toffoli gate constitute a universal gate, and propose an $(n,n)$ threshold quantum secret sharing scheme for any positive integer $n$. Then we use the quantum circuit composed of discrete Clifford group gates and Toffoli gate to evaluate the security.
\section{Quantum circuit and its related conclusions}
\label{sec:1}
In this section,we introduce a quantum circuit (See Fig.1, Fig.2) with a proof of security and give corresponding conclusions of the quantum circuit.
\par In Fig.1, the first $s$ qubits of the first column qubits are the quantum state secrets to be shared, and the later $t$ qubits are the ancilla to help implement the Toffoli gate. The notation $a_{x,y}$ labels the qubit on the $x$-th row and the $y$-th column, and $R_x$ labels the qubits in the $x$-th row, where $x\in\{1,2,\cdot\cdot\cdot,s\},y\in\{1,2,\cdot\cdot\cdot,n+1\}.$
\begin{figure}[H]
\centering
\includegraphics[scale=0.6]{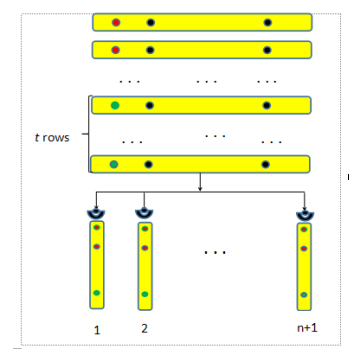}
\caption{Secret sharing process corresponding to $n+1$  participants}
\end{figure}
In Fig.2, $U$ represents the unitary matrix on the $x$-th row of qubits $R_x$ in Fig.1. When $i\in\{1,2,\cdot\cdot\cdot,n\},$  $V_i$  represents the corresponding unitary matrix of the $i$ column, and we order $A=V_nV_{n-1}\cdot\cdot\cdot V_1$. $W_i$  represents the corresponding unitary matrix of the $n+i$ column, and we order $B=W_nW_{n-1}\cdot\cdot\cdot W_1$, then $U=BA$.
\begin{figure}[H]
\centering
\includegraphics[scale=0.6]{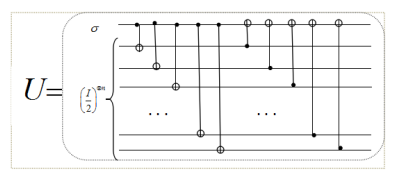}
\caption{ Schematic diagram of the corresponding circuit of  $n+1$  participants}
\end{figure}
\par According the quantum circuit given in Fig.1 and Fig.2, we give the following two properties.\\
\par \textbf{Property 1} when $n$ is an odd number,
$$U(\sigma\otimes I^{\otimes n-1})U^\dag=\sigma^{\otimes n}, \eqno{(1)}$$
where $\sigma\in P=\{I,X,Y,Z\}.$
\par \textbf{Prove}  For $\sigma=I,X,Y,Z,$  we give a proof of (1) respectively. That is, we prove the following equations are set up
$$U(I\otimes I^{\otimes n-1})U^\dag=I^{\otimes n}, \eqno{(2)}$$
$$U(X\otimes I^{\otimes n-1})U^\dag=X^{\otimes n},\eqno{(3)}$$
$$U(Y\otimes I^{\otimes n-1})U^\dag=Y^{\otimes n}, \eqno{(4)}$$
$$U(Z\otimes I^{\otimes n-1})U^\dag=Z^{\otimes n}, \eqno{(5)}$$
\par Firstly, by $U=BA$, it is easy to know that the above Eq.(2) is clearly established when $\sigma=I$ . Next, the  cases of $\sigma=\{X,Y,Z\}$ are proved by mathematical induction.
\par
\par When $\sigma=X$, we prove Eq.(3) to be established. First,the following Eq.(6) is established.
$$A(X\otimes I^{\otimes n-1})A^\dag=X^{\otimes n},n\in Z^+.  \eqno{(6)}$$
where~~$ A=|0\rangle\langle0|\otimes I^{\otimes n-1}+|1\rangle\langle1|\otimes X^{\otimes n-1}   $
\begin{equation*}
 \begin{split}
&~ ~~~~A(X\otimes I^{\otimes n-1})A^\dag\\
   &= (|0\rangle\langle0|\otimes I^{\otimes n-1}+|1\rangle\langle1|\otimes X^{\bigotimes n-1})(X\otimes I^{\otimes n-1})\\&~~~~(|0\rangle\langle0|\otimes I^{\otimes n-1}+|1\rangle\langle1|\otimes X^{\otimes n-1})     \\
   &= |0\rangle\langle1|\otimes I^{\otimes n-1}+|1\rangle\langle0|\otimes X^{\otimes n-1}= X^{\otimes n}.
 \end{split}
\end{equation*}
It is easy to know that when $n=3,5,$  Eq.(3) is held, so we assumed that Eq.(3) is established when $n=2k+1$, next certificate that Eq.(3) is also established when $n=2k+3$. According hypothesis and Eq.(6) we know that
\begin{center}
$B=W_{2k} W_{2k-1}\cdots W_1 ,$
\end{center}
\begin{center}
$U(X\otimes I^{\otimes2k})U^\dag=X^{\otimes 2k+1},$
\end{center}
\begin{center}
$A^\prime(X\otimes I^{\otimes 2k+2})(A^\prime)^\dag=X^{\otimes 2k+3}.$
\end{center}
When $n=2k+3,$
\begin{equation*}
 \begin{split}
&~ ~~~~B^\prime=W_{2k+2}^\prime W_{2k+1}^\prime W_{2k}^\prime W_{2k-1}^\prime\cdots W_1^\prime\\
   &=W_{2k+2}^\prime W_{2k+1}^\prime(W_{2k}\otimes I\otimes I)\cdots(W_1\otimes I\otimes I)\\
   &=W_{2k+2}^\prime W_{2k+1}^\prime(W_{2k}\otimes\cdots \otimes W_1\otimes I\otimes I)\\
   &=W_{2k+2}^\prime W_{2k+1}^\prime(B\otimes I\otimes I),
 \end{split}
\end{equation*}
where
\begin{center}
  $W_{2k+2}^\prime=I\otimes I^{\otimes 2k}\otimes|0\rangle\langle0|\otimes I+X\otimes I^{\otimes 2k}\otimes|1\rangle\langle1|\otimes I,$
  $W_{2k+1}^\prime=I\otimes I^{\otimes 2k}\otimes I\otimes|0\rangle\langle0|+X\otimes I^{\otimes 2k}\otimes I\otimes|1\rangle\langle1|.$
\end{center}
then
\begin{equation*}
 \begin{split}
      &~~~~ U^\prime(X\otimes I^{\otimes 2k+2})(U^\prime)^\dag      \\
   &  =B^\prime A^\prime(X\otimes I^{\otimes 2k+2})(A^\prime)^\dag (B^\prime)^\dag     \\
   &  =B^\prime(X^{\otimes 2k+3})(B^\prime)^\dag          \\
   &  =W_{2k+2}^\prime W_{2k+1}^\prime(X^{\otimes 2k+3})(W_{2k+1}^\prime)^\dag(W_{2k+2}^\prime)^\dag  \\
   &  =X^{\otimes 2k+1}\otimes(|0\rangle\langle1|\otimes X+|1\rangle\langle0|\otimes X)      \\
   &  =X^{\otimes 2k+3}.
 \end{split}
\end{equation*}
It is known that the Eq.(3) holds according the inductive hypothesis.
\par When $\sigma=Y$, we prove Eq.(4) to be established. Firstly, we suppose $Y=|1\rangle\langle0|-|0\rangle\langle1|$, and it is clear from  $Y=i|1\rangle\langle0|-i|0\rangle\langle1|$  that it is also true. The same reason has the next form
$$A(Y\otimes I^{\otimes n-1})A^\dag=Y\otimes X^{\otimes n-1},n\in Z^+.  \eqno{(7)}$$
It is easy to know that when $n=3,5,$  Eq.(4) is held, so we assumed that Eq.(4) is established when $n=2k+1$, next certificate that Eq.(4) is also established when $n=2k+3$. According hypothesis and Eq.(7) wo know that
\begin{center}
$B=W_{2k} W_{2k-1}\cdots W_1, $
\end{center}
\begin{center}
$U(Y\otimes I^{\otimes 2k})U^\dag=Y^{\otimes 2k+1},$
\end{center}
\begin{center}
$A^\prime(Y\otimes I^{\otimes 2k+2})(A^\prime)^\dag=Y\otimes X^{\otimes 2k+2}.$
\end{center}
When $n=2k+3,$
\begin{equation*}
 \begin{split}
      & ~ ~~~~U^\prime(Y\otimes I^{\otimes 2k+2})(U^\prime)^\dag      \\
   &  =B^\prime(Y\otimes X^{\otimes 2k+2})(B^\prime)^\dag           \\
   &  =W_{2k+2}^\prime W_{2k+1}^\prime(Y^{\otimes 2k+1}\otimes X\otimes X)(W_{2k+1}^\prime)^\dag(W_{2k+2}^\prime)^\dag  \\
   &  =Y^{\otimes 2k+1}\otimes(-|0\rangle\langle1|\otimes Y+|1\rangle\langle0|\otimes Y)      \\
   &  =Y^{\otimes 2k+3}.
 \end{split}
\end{equation*}
It is known that the Eq.(4) holds according the inductive hypothesis.
\par When $\sigma=Z,$ we prove Eq.(5) to be established. The same reason has the next form
$$A(Z\otimes I^{\otimes n-1})A^\dag=Z\otimes X^{\otimes n-1},n\in Z^+.  \eqno{(8)}$$
It is easy to know that when $n=3,5,$  Eq.(5) is held, so assumed that Eq.(5) is established when $n=2k+1$, next certificate that Eq.(5) is also established when $n=2k+3$. According hypothesis and Eq.(8) we know that
\begin{center}
$B=W_{2k} W_{2k-1}\cdots W_1, $
\end{center}
\begin{center}
$U(Z\otimes I^{\otimes 2k})U^\dag=Z^{\otimes 2k+1},$
\end{center}
\begin{center}
$A^\prime(Z\otimes I^{\otimes 2k+2})(A^\prime)^\dag=Z\otimes I^{\otimes 2k+2}.$
\end{center}
When $n=2k+3,$
\begin{equation*}
 \begin{split}
      &~ ~~~~ U^\prime(Z\otimes I^{\otimes 2k+2})(U^\prime)^\dag      \\
   &  =B^\prime(Z\otimes I^{\otimes 2k+2})(B^\prime)^\dag          \\
   &  =W_{2k+2}^\prime W_{2k+1}^\prime(Z^{\otimes 2k+1}\otimes I\otimes I)(W_{2k+1}^\prime)^\dag(W_{2k+2}^\prime)^\dag  \\
   &  =Z^{\otimes 2k+1}\otimes(|0\rangle\langle0|\otimes Z-|1\rangle\langle1|\otimes Z)      \\
   &  =Z^{\otimes 2k+3}.
 \end{split}
\end{equation*}
It is known that the Eq.(5) holds according the inductive hypothesis.
\\To sum up: the Property 1 has to be proved.
\par \textbf{Property 2}~~when $n$ is an even number,
$$U(\sigma\otimes I^{\otimes n-1})U^\dag=\left\{
\begin{array}{lll}
&I\otimes\sigma^{\otimes n-1},  ~  \sigma\in\{I,X\}\\
& Z\otimes\sigma^{\otimes n-1},   \sigma\in\{Y,Z\}.
\end{array}
\right.  \eqno{(9)}$$
\par \textbf{Prove}  For $\sigma=I,X,Y,Z,$  we give a proof of (9) respectively. That is, we prove the following equations are set up
$$U(I\otimes I^{\otimes n-1})U^\dag=I^{\otimes n}.  \eqno{(10)}$$
$$U(X\otimes I^{\otimes n-1})U^\dag=I\otimes X^{\otimes n-1} .  \eqno{(11)}$$
$$U(Y\otimes I^{\otimes n-1})U^\dag=Z\otimes Y^{\otimes n-1}. \eqno{(12)}$$
$$ U(Z\otimes I^{\otimes n-1})U^\dag=Z\otimes Z^{\otimes n}.  \eqno{(13)}$$
\par Firstly, by $U=BA$, it is easy to know that the above Eq.(10) is clearly established when $\sigma=I$ . Next, the  cases of $\sigma=\{X,Y,Z\}$ are proved by mathematical induction.
\par When $\sigma=X$, it is easy to know that when $n=2,4,$ Eq.(11) is held, so we assumed that Eq.(11) is established when $n=2k$, next certificate that Eq.(11) is also established when $n=2k+2$. According hypothesis and Eq.(6) we know that
\begin{center}
$B=W_{2k-1} W_{2k-2}\cdots W_1 ,$
\end{center}
\begin{center}
$U(X\otimes I^{\otimes 2k-1})U^\dag=I\otimes X^{\otimes 2k-1},$
\end{center}
\begin{center}
$A^\prime(X\otimes I^{\otimes 2k+1})(A^\prime)^\dag=X^{\otimes 2k+2},$
\end{center}
when $n=2k+2,$
\begin{equation*}
 \begin{split}
&~ ~~~~B^\prime=W_{2k+1}^\prime W_{2k}^\prime W_{2k-1}^\prime W_{2k-2}^\prime\cdots W_1^\prime\\
   &=W_{2k+1}^\prime W_{2k}^\prime(W_{2k-1}\otimes I\otimes I)\cdots(W_1\otimes I\otimes I)\\
   &=W_{2k+1}^\prime W_{2k}^\prime(W_{2k-1}\otimes\cdots \otimes W_1\otimes I\otimes I)\\
   &=W_{2k+1}^\prime W_{2k}^\prime(B\otimes I\otimes I),
 \end{split}
\end{equation*}
where
\begin{center}
  $W_{2k}^\prime=I^{\otimes 2k}\otimes|0\rangle\langle0|\otimes I+X\otimes I^{\otimes 2k-1}\otimes|1\rangle\langle1|\otimes I,$
  $W_{2k+1}^\prime=I^{\otimes 2k}\otimes I\otimes|0\rangle\langle0|+X\otimes I^{\otimes 2k-1}\otimes I\otimes|1\rangle\langle1|,$
\end{center}
then
\begin{equation*}
 \begin{split}
      & ~~~ ~U^\prime(X\otimes I^{\otimes 2k+1})(U^\prime)^\dag      \\
   &  =B^\prime A^\prime(X\otimes I^{\otimes 2k+1})(A^\prime)^\dag(B^\prime)^\dag     \\
   &  =B^\prime(X^{\otimes 2k}\otimes X^{\otimes 2})(B^\prime)^\dag          \\
   &  =W_{2k+1}^\prime W_{2k}^\prime(I\otimes X^{\otimes 2k+1})(W_{2k}^\prime)^\dag(W_{2k+1}^\prime)^\dag  \\
   &  =I\otimes X^{\otimes 2k+1}.
 \end{split}
\end{equation*}
It is known that the Eq.(11) holds according the inductive hypothesis.
\par When $\sigma=Y$, it is easy to know that when $n=2,4,$ Eq.(12) is held, so we assumed that Eq.(12) is established when  $n=2k$, next certificate that Eq.(12) is also established when $n=2k+2$ . According hypothesis and Eq.(7) we know that
\begin{center}
$B=W_{2k-1} W_{2k-2}\cdots W_1 ,$
\end{center}
\begin{center}
$U(Y\otimes I^{\otimes 2k-1})U^\dag=Z\otimes Y^{\otimes 2k-1},$
\end{center}
\begin{center}
$A^\prime(Y\otimes I^{\otimes 2k+1})(A^\prime)^\dag=Y\otimes X^{\otimes 2k+1},$
\end{center}
when $n=2k+2,$
\begin{center}
  $B^\prime=W_{2k+1}^\prime W_{2k}^\prime(B\otimes I\otimes I),$
\end{center}
then
\begin{equation*}
 \begin{split}
      & ~~~ ~U^\prime(Y\otimes I^{\otimes 2k+1})(U^\prime)^\dag      \\
   &  =B^\prime A^\prime(X\otimes I^{\otimes 2k+1})(A^\prime)^\dag(B^\prime)^\dag     \\
   &  =B^\prime(Y\otimes X^{\otimes 2k+1})(B^\prime)^\dag          \\
   &  =W_{2k+1}^\prime W_{2k}^\prime(Z\otimes Y^{\otimes 2k-1}\otimes X^{\otimes 2})(W_{2k}^\prime)^\dag(W_{2k+1}^\prime)^\dag  \\
   &  =Z\otimes Y^{\otimes 2k+1}.
 \end{split}
\end{equation*}
It is known that the Eq.(12) holds according the inductive hypothesis.
\par When $\sigma=Z$,  it is easy to know that when $n=2,4,$ Eq.(12) is held, so we assumed that Eq.(13) is established when $n=2k$, next certificate that Eq.(13) is also established when $n=2k+2$ . According hypothesis and Eq.(8) we know that
\begin{center}
$B=W_{2k-1} W_{2k-2}\cdots W_1 ,$
\end{center}
\begin{center}
$U(Z\otimes I^{\otimes 2k-1})U^\dag=Z^{\otimes 2k},$
\end{center}
\begin{center}
$A^\prime(Z\otimes I^{\otimes 2k+1})(A^\prime)^\dag=Z\otimes I^{\otimes 2k+1},$
\end{center}
when $n=2k+2,$
\begin{center}
  $B^\prime=W_{2k+1}^\prime W_{2k}^\prime(B\otimes I\otimes I),$
\end{center}
then
\begin{equation*}
 \begin{split}
      & ~~~ ~U^\prime(Z\otimes I^{\otimes 2k+1})(U^\prime)^\dag      \\
   &  =B^\prime A^\prime(Z\otimes I^{\otimes 2k+1})(A^\prime)^\dag(B^\prime)^\dag     \\
   &  =B^\prime(Z\otimes I^{\otimes 2k-1}\otimes I^{\otimes 2})(B^\prime)^\dag          \\
   &  =W_{2k+1}^\prime W_{2k}^\prime(Z^{\otimes 2k}\otimes I^{\otimes 2})(W_{2k}^\prime)^\dag(W_{2k+1}^\prime)^\dag  \\
   &  =Z^{\otimes 2k+2}.
 \end{split}
\end{equation*}
It is known that the Eq.(13) holds according the inductive hypothesis.
\\To sum up, the Property 2 is proved.
\section{Secret sharing scheme}
\par In the Ref.[14], an $(n,n)$ threshold quantum secret sharing scheme is proposed, in which the number of participants is limited to $n=4k+1,k\in Z^+$, and the scheme use quantum circuits to evaluate security after the participants get their own quantum states.  In this section, we discuss the number of participants $n$ in any case ( $n\in Z^+$ ).  The specific scheme is as follows:
\par \textbf{Input procedure}
\par The first $s$  bits in the first column are initialized as a quantum secret, and the later $t$  bits are auxiliary state: $|\phi_+\rangle=U_T(H_1\otimes H_2)|000\rangle=\frac{1}{2}(|000\rangle+|010\rangle+|100\rangle+|111\rangle),$ let $s=3k,t=3k^\prime, k^\prime/k\in Z^+.$ The rest of the column is initialized to the maximum mixed state $\frac{I}{2}$,   where $I,X,Y,Z$ are common Pauli operator, and $U_T$  is a Toffoli gate.
\par \textbf{Encoding procedure}
\par When $x\in\{1,2,\cdots,s\}$, $U$ acts on $x$-th qubit $R_x$, so  $U^{\otimes s}$  act on $s(n+1)$  bit quantum state and encodes the quantum state secret into a highly mixed state.
\par \textbf{Sharing procedure}
\par Prepared $U$  encrypted  $n+1$ column mixed quantum state, Alice holding first column quantum state, $y$-th column quantum state send to the $y-1$  participant, where $y\in\{2,3,\cdots,n+1\}.$
\par \textbf{Decoding procedure}
\par {(a) Collecting the shared state of $n+1$  participants;}
\par {(b) The unitary matrix  $U^\dag$  acts on the mixed quantum state shared by the $n+1$ column, discards the rest of the columns, and the remaining first column $s$  bits are quantum state secrets.}
\section{Scheme analysis}
\par In order to analyze quantum circuits acting on shared secrets, each participant performs quantum computation on his own share only, and its computation is carried out between sharing and decryption procedure. In Ref. [27], it is shown that any gate in $C_3\backslash C_2$ and Clifford group gates can form a set of universal gate. In this paper, we consider a set of discrete gate sets consisting of Clifford group gates and Toffoli gate. where $U_T=(|00\rangle\langle00|+|01\rangle\langle01|+|10\rangle\langle10|)\otimes I+|11\rangle\langle11|\otimes X.$  Quantum circuit consists of any multiple Clifford gates and a constant $k^\prime/k$  Toffoli gates. Now we consider a sequence gate  $\mathcal{U}=(U_1,\cdots,U_L)$  that acts on the  $s$ bit quantum state secret, where  $U_1,\cdots,U_L$ gates are all unitary and are known to the participants.
\par When $U_i$ is a Clifford gate, each participant implements  $U_i$ on a list of its own qubit subsets, so  $n+1$ party participants jointly implement $U_i^{\otimes n+1}$. When $U_i$  is a single-qubit Clifford gates, as shown in Fig.3, when $U_i$  is a double-qubit CNOT gate, as shown in Fig.4.
\begin{figure}[htbp]
\begin{minipage}[t]{0.35\linewidth}
\centering
\includegraphics[height=2.5cm,width=5.0cm]{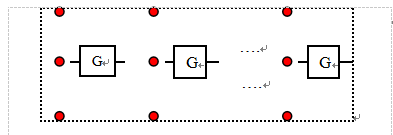}
\caption{Schematic diagram of Clifford gate action}
\end{minipage}%
\hfill
\begin{minipage}[t]{0.5\linewidth}
\centering
\includegraphics[height=2.5cm,width=5.0cm]{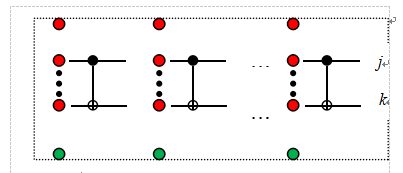}
\caption{Schematic diagram of CONT gate action}
\end{minipage}
\end{figure}
\par When $U_i$ is a Toffoli gate, each participant can perform a constant  $k^\prime/k$ Toffoli gates on its own quantum secret. To implement the $j$-th  Toffoli gate at the $m,n,l$  bit qubit, each participant operates like Fig.5, where $m,n,l\in\{1,\cdots,s\},j\in\{1,\cdots,k^\prime/k\}.$
\begin{figure}[H]
\centering
\includegraphics[scale=0.6]{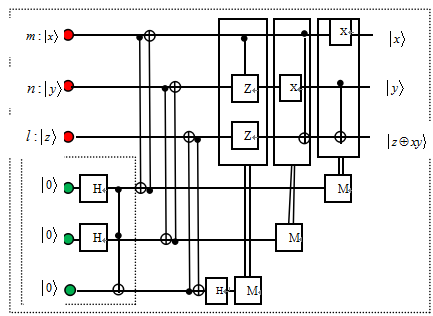}
\caption{ Toffoli gate diagram}
\end{figure}
\par Here the $M$  frame represents the measuring device, when the result is classic bit 1, the operation in the above box is implemented, and the Toffoli gate is finally realized. Here $|x\rangle,|y\rangle,|z\rangle$  is the quantum state of the  $m,n,l$ bit, $|\phi_+\rangle=U_T(H_1\otimes H_2)|000\rangle=\frac{1}{2}(|000\rangle+|010\rangle+|100\rangle+|111\rangle)$ is an auxiliary particle. From Fig.5, it can be seen that each implementation of a Toffoli gate consumes an auxiliary particle $|\phi_+\rangle.$
\par In order to better understand the above scheme and analyze the implementation of quantum gates on the shared quantum secret, we give two examples of $n=4,5.$
\par \textbf{Example} Suppose that the shared quantum state secret is  $\sigma=X\otimes Y\otimes Z,t=6.$ After the action of the unitary matrix $U$ in the scheme, the final state is as follows
\begin{center}
 $ \tilde{\rho}_{sec}=(X\otimes Y\otimes Z)^{\otimes 5}\otimes\Phi,$
\end{center}
 $\Phi=U(|\phi_+\rangle\langle\phi_+|\otimes(I^{\otimes 3})^{\otimes 4})U^\dag$  is the result state of the auxiliary particle after the action of the unitary matrix $U$. Assuming that the first three participants (excluding Alice) are dishonest, and the fourth participant are honest, they trace the quantum state held by fourth participant, namely, find the reduced density matrix corresponding to the four dimensional subsystems including Alice
\begin{equation*}
 \begin{split}
      & ~~~  ~~~\mathrm{tr}_5(\tilde{\rho}_{sec})=\mathrm{tr}_5((X\otimes Y\otimes Z)^{\otimes 5}\otimes\Phi)   \\
   &  =(X\otimes Y\otimes Z)^{\otimes 4}\otimes\Phi^\prime\cdot \mathrm{tr}(X\otimes Y\otimes Z\otimes\Phi_5)     \\
   &  =(X\otimes Y\otimes Z)^{\otimes 4}\otimes\Phi^\prime\cdot \mathrm{tr}(X\otimes Y\otimes Z)\cdot \mathrm{tr}(\Phi_5)          \\
   &  =(X\otimes Y\otimes Z)^{\otimes 4}\otimes\Phi^\prime\cdot 0\cdot \mathrm{tr}(\Phi_5)   \\
   &  =0,
 \end{split}
\end{equation*}
$\Phi^\prime$  is the first four column entangled states of $\Phi$ , and  $\Phi_5$ is the fifth column quantum state of  $\Phi$. The rest of the three participants and the Alice did not get any information about the quantum secret. We consider a sequence gates $\mathcal{U}=(C_1,C_2,U_{T_1},C_3,U_{T_2},C_4)$ on the $s=3$  bits quantum state secret, where $C_1,C_2,C_3,C_4$ are Clifford gates, $U_{T_1}$ and $U_{T_2}$ are Toffoli gates, frame  $U_{T_1}$ and  $U_{T_2}$ in the diagram as in Fig.5; Each unitary gate is known to the participants.  The operations performed by each participant are shown in Fig.6.
\begin{figure}[H]
\centering
\includegraphics[scale=0.6]{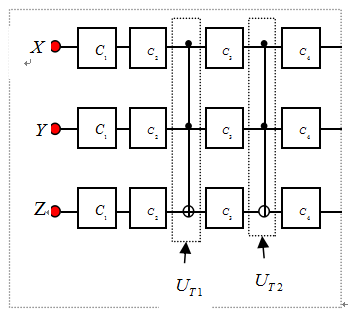}
\caption{ Quantum diagram of a sequence gates }
\end{figure}
\par \textbf{Example} Suppose that the shared quantum state secret is  $\sigma=X\otimes Y\otimes Z,t=6.$ After the action of the unitary matrix $U$ in the scheme, the final state is as follows
\begin{center}
 $ \tilde{\rho}_{sec}=(I\otimes Z\otimes Z)\otimes\sigma^{\otimes 5}\otimes\Phi,$
\end{center}
$\Phi=U(|\phi_+\rangle\langle\phi_+|\otimes(I^{\otimes 3})^{\otimes 4})U^\dag$ is the result state of the auxiliary particle after the action of the unitary matrix. Assuming that the first four participants (excluding Alice) try to get secret information, then they trace the quantum state held by fourth participant, namely, find the reduced density matrix corresponding to the five dimensional subsystems including Alice
\begin{equation*}
 \begin{split}
      & ~~~  ~~~\mathrm{tr}_6(\tilde{\rho}_{sec})=\mathrm{tr}_6((I\otimes Z\otimes Z)\otimes\sigma^{\otimes 5}\otimes\Phi)   \\
   &  =(I\otimes Z\otimes Z)\otimes\sigma^{\otimes 4}\otimes\Phi^{\prime\prime}\cdot \mathrm{tr}(X\otimes Y\otimes Z\otimes\Phi_5^\prime)     \\
   &  =(I\otimes Z\otimes Z)\otimes\sigma^{\otimes 4}\otimes\Phi^{\prime\prime}\cdot 0\cdot \mathrm{tr}(\Phi_5)          \\
   &  =0.
 \end{split}
\end{equation*}
$\Phi^{\prime\prime}$  is the first four column entangled states of $\Phi$ , and  $\Phi_5^\prime$ is the sixth column quantum state of  $\Phi$. The rest of the four participants and the Alice did not get any information about the quantum secret. In the same way, we consider a sequence gates $\mathcal{U}=(C_1,C_2,U_{T_1},C_3,U_{T_2},C_4)$ on the $s=3$  bits quantum state secret,  each unitary gate is known to the participants.  The operations performed by each participant are shown in Fig.6.
\section{Security discussion}
\par ~~ ~~\textbf{Dishonest participants attack:} A  $(k,n)$-threshold quantum secret sharing scheme satisfies two properties: (1) any  $k$ or more parties can perfectly reconstruct the secret quantum state, (2) any $k-1$  or fewer parties can collectively deduce no information at all about the secret quantum state. In this paper, when $k=n$ the first property  is clearly hold, it can be seen that the coding program is completely reversible in the encryption phase. Next, we discuss the second property. The secret quantum state before encrypting is
$$\rho_{sec}=2^{-s}\displaystyle{\sum_{\sigma\in P^{\otimes s}}}\omega_\sigma\sigma , \eqno{(14)}$$
where $\sigma=\sigma_1\otimes\sigma_2\otimes\cdots\otimes\sigma_s,P=\{I,X,Y,Z\};$  it is coefficient $\omega_\sigma$ for the non-trivial Pauli oprators  $\sigma$ in  $P^{\otimes s}$, and $\omega_\sigma=1$   when  $\sigma$ is the trivial Pauli oprator.
\par When $n+1$  is odd, it is known by the Property 1 that the final state of the encrypted post is£º
$$\tilde{\rho}_{sec}=2^{-s}(\displaystyle{\sum_{\sigma\in P^{\otimes s}}}\omega_\sigma\sigma^{\otimes n+1})\otimes\Phi .  \eqno{(15)}$$
 $\Phi=U(|\phi_+\rangle\langle\phi_+|\otimes(I^{\otimes3})^{\otimes n})U^\dag.$  Assuming that the  $y$-th  participant is honest, the other $n-1$  participants can not get any information about the quantum secret, because we can get the reduced density matrix on the $n$  dimension subsystem by tracing the quantum state held by the  $y$-th participant
\begin{equation*}
 \begin{split}
      & ~~~  ~~~\tilde{\rho}_{sec}=2^{-s}(\displaystyle{\sum_{\sigma\in P^{\otimes s}}}\omega_\sigma\sigma^{\otimes n+1})\otimes\Phi    \\
   &  =2^{-s}((I^{\otimes s})^{\otimes n+1}+\displaystyle{\sum_{\sigma\in P^{\otimes s}\setminus I^{\otimes s}}}\omega_\sigma\sigma^{\otimes n+1})\otimes\Phi,     \\
 \end{split}
\end{equation*}
where\\
\begin{center}
  $\mathrm{tr}_{y+1}(\displaystyle{\sum_{\sigma\in P^{\otimes s}\setminus I^{\otimes s}}}\omega_\sigma\sigma^{\otimes n+1})=\displaystyle{\sum_{\sigma\in P^{\otimes s}\setminus I^{\otimes s}}}\omega_\sigma \mathrm{tr}_{y+1}(\sigma^{\otimes n+1})=0,$
\end{center}
then
\begin{equation*}
 \begin{split}
      & ~~~  ~~~\mathrm{tr}_{y+1}\{2^{-s}((I^{\otimes s})^{\otimes n+1}+\displaystyle{\sum_{\sigma\in P^{\otimes s}\setminus I^{\otimes s}}}\omega_\sigma\sigma^{\otimes n+1})\}   \\
   &  =2^{-s}\{\mathrm{tr}_{y+1}(I^{\otimes s})^{\otimes n+1}+0\}  \\
   &  =(I^{\otimes s})^{\otimes n}.
 \end{split}
\end{equation*}
\par When $n+1$  is even, it is known by the Property 2 that the final state of the encrypted is£º
\begin{center}
  $\tilde{\rho}_{sec}=2^{-s}(\displaystyle{\sum_{\sigma\in P^{\otimes s}}}\omega_\sigma\theta\otimes\sigma^{\otimes n})\otimes\Phi.     $
\end{center}
$\Phi=U(|\phi_+\rangle\langle\phi_+|\otimes(I^{\otimes3})^{\otimes n})U^\dag.$  Where $\theta$  is the product state of the first column of quantum bits held by Alice. Similarly, assuming that the $y$-th  participant is honest, the other $n-1$  participants can not get any information about the quantum secret, because we can get the reduced density matrix on the $n$  dimension subsystem by tracing the quantum state held by the $y$-th participant
\begin{equation*}
 \begin{split}
      & ~~~  ~~~\tilde{\rho}_{sec}=2^{-s}(\displaystyle{\sum_{\sigma\in P^{\otimes s}}}\omega_\sigma\theta\otimes\sigma^{\otimes n})\otimes\Phi    \\
   &  =2^{-s}(\theta\otimes(I^{\otimes s})^{\otimes n}+\displaystyle{\sum_{\sigma\in P^{\otimes s}\setminus I^{\otimes s}}}\omega_\sigma\theta\otimes\sigma^{\otimes n})\otimes\Phi.     \\
 \end{split}
\end{equation*}
Where
\begin{center}
  $\mathrm{tr}_{y+1}(\displaystyle{\sum_{\sigma\in P^{\otimes s}\setminus I^{\otimes s}}}\omega_\sigma\theta\otimes\sigma^{\otimes n})=\displaystyle{\sum_{\sigma\in P^{\otimes s}\setminus I^{\otimes s}}}\omega_\sigma\theta\otimes \mathrm{tr}_{y}(\sigma^{\otimes n})=0,$
\end{center}
then
\begin{equation*}
 \begin{split}
      & ~~~  ~~~\mathrm{tr}_{y+1}\{2^{-s}(\displaystyle{\sum_{\sigma\in P^{\otimes s}}}\omega_\sigma\theta\otimes\sigma^{\otimes n})\}\\
  &  =\mathrm{tr}_{y+1}\{2^{-s}(\theta\otimes(I^{\otimes s})^{\otimes n}+\displaystyle{\sum_{\sigma\in P^{\otimes s}\setminus I^{\otimes s}}}\omega_\sigma\theta\otimes\sigma^{\otimes n})\}   \\
   &  =2^{-s}\{\theta\otimes \mathrm{tr}_{y}(I^{\otimes s})^{\otimes n}+\displaystyle{\sum_{\sigma\in P^{\otimes s}\setminus I^{\otimes s}}}\omega_\sigma\theta\otimes \mathrm{tr}_y(\sigma^{\otimes n})\}  \\
   &  =\theta\otimes(I^{\otimes s})^{\otimes n-1}.
 \end{split}
\end{equation*}
In the two cases, the dishonest participant can only get a highly mixed state, so there is no information about the quantum secret.
\par Second property is also satisfied when the set of discrete gates composed of Clifford group gates and Toffoli gate act on shared quantum state secret. Suppose that only one of the $n$  participants is honest, while the remaining  $n-1$ participants can perform any operation, which proves that the result of the honest party's announcement is uniformly random and independent of the behavior of other participants [14]. According to the effect of the gate evaluation, after the evaluation of the $i$-th gate the state of the system has the form
$$\rho_i=\sum b_i\psi_{Alice}\otimes(\frac{\sigma\otimes\gamma}{2^N})\otimes\chi_i .  \eqno{(16)}$$
where $\sigma\in P^{\otimes s},\gamma$ is an auxiliary particle that is not destroyed, $b_i$ is a set of ccalars, $\psi_{Alice}$ is a column of quantum states held by Alice, and  $\chi_i$  is a set of operators on the dishonest parties¡¯ system. Honesty and other systems exist in the form of the product state, therefore, the results of the honest measure do not convey any useful information .
\section{ Conclusions}
\label{sec:2}
 In this paper, we have constructed an $(n,n)$ threshold quantum secret sharing scheme for $n$ with arbitrary number ($n\in Z^+$) of participants, which is based on a quantum circuit composed of Clifford group gates and Toffoli gate. Because of the universal quantum logic gate has different forms, looking for general quantum gate model set and consider other types of universal quantum circuit gate construction, and more secure quantum secret sharing scheme based on these quantum circuits is a problem to be studied in the future.

\end{document}